\documentclass[twocolumn,showpacs,preprintnumbers,amsmath,amssymb]{revtex4}

\usepackage{epsf}
\usepackage{graphicx}
\usepackage{dcolumn}
\usepackage{bm}
\def\be{\begin{equation}}
\def\ee{\end{equation}}
\def\bea{\begin{eqnarray}}
\def\eea{\end{eqnarray}}

\newlength{\dinwidth}
\newlength{\dinmargin}
\def\fun#1#2{\lower3.6pt\vbox{\baselineskip0pt\lineskip.9pt
  \ialign{$\mathsurround=0pt#1\hfil##\hfil$\crcr#2\crcr\sim\crcr}}}

\begin{document}
 \tighten
\vskip 3cm

\

\title{\Large\bf Can we have  inflation with $\Omega > 1$?}

\author{\bf Andrei Linde}\thanks{Extended version of the talk given at the conference COSMO-01, September 4, 2001, Rovaniemi, Finland}

\affiliation{ {Department
  of Physics, Stanford University, Stanford, CA 94305-4060,
USA}    }

\date{March 11, 2003 \\ \phantom{}}
 
 {
\begin{abstract}
It is very difficult to obtain a realistic model of a closed inflationary
universe. Even if one fine-tunes the total number of e-folds to be
sufficiently small, the resulting universe typically
has ${\delta\rho\over \rho} \sim {\Delta T \over T} = O(1)$ on the scale of
the horizon. We describe a class of models where this problem can be
resolved. The models are unattractive and fine-tuned, so the flatness of
the universe remains a generic prediction of inflationary cosmology.
Nevertheless one should keep in mind that with the fine-tuning at the level
of about one percent one can obtain a semi-realistic model  of an inflationary
universe with $\Omega > 1$. The spectrum of density perturbations in this model 
may have a cut-off on the scale of the horizon. Similar approach may be valid in 
application to a compact inflationary universe with a nontrivial topology.
\end{abstract}}

\maketitle

 \parskip 2pt

\section{Introduction}
Flatness of the universe is one of the most robust predictions of
inflationary cosmology. Similarly, nearly all inflationary models predict density perturbations with a flat spectrum. Both predictions are in an excellent agreement with recent observations, $\Omega = 1.02\pm 0.02$, and $n_s= 0.97\pm 0.03$, see e.g. \cite{Spergel:2003cb} - \cite{Contaldi}. With the data like that, it would seem unnecessary to return to investigation of  more complicated models of inflation, with $\Omega \not = 1$ and with the spectrum having various `features.'

However, there is only one way to prove that the flatness of the universe and the flatness of the spectrum are indeed reliable and robust predictions of inflation: We must  check whether it is really difficult to violate these rules. Here we will return to the discussion of the possibility to have inflationary models with $\Omega \not = 1$. 

Despite many interesting attempts, we still do not have
any realistic models of an open inflationary universe satisfying the observational constraint $1-\Omega\ll 1$.  The only available semi-realistic model of open inflation with $1-\Omega\ll 1$ is rather ugly; it requires a fine-tuned potential of a very peculiar shape  \cite{Linde:1998iw}.  But maybe we can be more lucky with a closed universe, or a compact flat universe with nontrivial topology? Recently this question resurfaced again in relation to WMAP results \cite{Tegmark:2003ve,Uzan:2003nk,Efstathiou}, even though these results are quite consistent with the simplest assumption that $\Omega = 1$. The only potential problem of the standard model of a flat universe with a flat spectrum is related to the suppression of the large-angle perturbations of CMB \cite{Spergel:2003cb} and possible correlations between lower multipoles \cite{Tegmark:2003ve}. However, taking into account cosmic variance, it is not quite clear yet whether these results are statistically significant. 
Nevertheless, it gives us a good excuse to reexamine possible inflationary models of a closed (or compact) universe. 

The only way to obtain inflationary universe with $\Omega > 1$ is to
assume that the universe inflated only by about $e^{60}$ times. Exact
value of the number of e-folds $N$ depends on details of reheating
and may somewhat differ from 60. Let us assume for definiteness that for
$N = 60$ one would have $\Omega = 1.1$.  Then one can show that for $N =
59.5$ and the same value of the Hubble constant one would have $\Omega
\approx 1.3$, whereas for $N = 60.5$ one would have $\Omega \approx
1.03$. Thus in order to obtain $\Omega = 1.1\pm 0.05$ one would need to have
$N = 60$ with accuracy of about $1\%$. Meanwhile, the typical number of
e-folds, say, in chaotic inflation scenario in the theory ${m^2\over
2} \phi^2$ is not 60 but rather $10^{12}$ \cite{book}.

One can construct models where inflation leads to expansion of the
universe by the factor $e^{60}$. However, if the universe does not
inflate long enough to become flat, then by the same token it does not
inflate long enough to become homogeneous and isotropic.  Moreover, in
most of such models small number of e-foldings simultaneously implies that
density perturbations on the horizon are extremely large.  Indeed, in most
inflationary models the process of inflation begins at the point where
density perturbations ${\delta\rho\over\rho} \sim {H^2\over \dot\phi}$
are very large. The simplest example is the new inflation scenario
\cite{New}, where inflation begins at the top of the effective potential
with $\dot\phi = 0$. If there are less than $60$ e-foldings from this
moment to the end of inflation, then we would see extremely large density
perturbations on the scale of the horizon. Similarly, in the simplest models of chaotic inflation \cite{Chaotic} the process of inflation is expected to begin at very large values of the scalar field, where the density perturbations are also very large. This is related to the observation that inflation usually is eternal \cite{Vilenkin:xq,Linde:fc}. The boundary of the regime of eternal inflation corresponds to ${\delta\rho\over\rho} = O(1)$ \cite{book}. In a small closed universe with $\Omega$ noticeably different from $1$ we would see these large perturbations on the horizon.

Thus, the main reason why it is so difficult to construct inflationary
models with $\Omega \not = 1$ is not the issue of fine tuning of the
parameters of the models, which is necessary to obtain the universe
inflating exactly $e^{60}$ times, but the problem of obtaining a
homogeneous universe after inflation.

For a long time it was not quite clear how one can resolve this problem; in  \cite{BGT} it was even argued that this is
impossible. A solution of this problem was proposed in 
\cite{Lab,Omega}. The
main idea is to use the well known fact that the region of space created in the
process of quantum tunneling tends to be spherically symmetric if the tunneling probability is very small. Such bubbles tend  to expand in a
spherically symmetric
way. Thus, if one
could associate the visible part of the universe with an interior of one
such region, one would solve the homogeneity, isotropy and horizon problems, and then all other
problems, such as the flatness problem, 
will be solved by the subsequent relatively short stage of inflation.

In order to make inflation short, 
one can consider, for example, a particular version of the chaotic inflation
scenario
\cite{Chaotic} with the effective potential
\begin{equation}\label{1}
V(\phi) = {m^2 \phi^2\over 2}\, \exp{\Bigl({\phi\over CM_{\rm P}}\Bigr)^2} \ .
\end{equation}
Here $M_P = 1/\sqrt G \sim 1.2 \times 10^{19}$ GeV.  Potentials of such type often appear in supergravity. In this theory
inflation occurs only in the interval ${M_{\rm P}\over 2} {\
\lower-1.2pt\vbox{\hbox{\rlap{$<$}\lower5pt\vbox{\hbox{$\sim$}}}}\ } \phi {\
\lower-1.2pt\vbox{\hbox{\rlap{$<$}\lower5pt\vbox{\hbox{$\sim$}}}}\ }
CM_{\rm P}$. The most natural way to realize inflationary scenario in this
theory is
to assume that the universe was created ``from nothing''  with the field $\phi$
in the interval  ${M_{\rm P}\over 2} {\
\lower-1.2pt\vbox{\hbox{\rlap{$<$}\lower5pt\vbox{\hbox{$\sim$}}}}\ } \phi {\
\lower-1.2pt\vbox{\hbox{\rlap{$<$}\lower5pt\vbox{\hbox{$\sim$}}}}\ }
CM_{\rm P}$. The universe at the moment of its creation has a size $H^{-1}$,
and then it begins inflating as $H^{-1} \cosh{Ht}$. According to
\cite{Creation} - \cite{Vilenkin}, the probability of creation of an inflationary
universe is
suppressed by
\begin{equation}\label{2}
P \sim \exp\Bigl(-{3M^4_{\rm P}\over 8V(\phi)}\Bigr) \ .
\end{equation}

Strictly speaking, this expression was derived only for the creation of de Sitter space with the constant vacuum energy $V(\phi)$ corresponding to an extremum of the effective potential. For the description of the quantum creation of inflationary universe on the slope of the potential one should use the `gondola' instantons found in   \cite{Bousso:1998ed}. These instantons are singular, but the singularity is nearly harmless since the analytical continuation occurs through the regular part of the instanton. The singularity gives a contribution to the tunneling action which is small for the tunneling going to the inflationary universe in the slow-roll regime. However, the boundary terms can be significant and tend to decrease the exponential suppression of the tunneling probability in the calculation of the tunneling to the steep part of the potential. This is a delicate issue which deserves separate investigation. However, we hope that this fact does not affect our qualitative conclusions because we are going to find a regime where the (relative) suppression of the tunneling will be inefficient, see next section.

The maximum of the probability of creation of an inflationary universe appears near the upper range of values
of the field $\phi$ for which inflation is possible, i.e. at $\phi_0 \sim C
M_{\rm P}$.
Then the size of the
newly born universe  expands by the factor $\exp({2\pi
\phi_0^2M_{\rm P}^{-2}})\sim \exp({2\pi C^2})$ during the
stage of inflation \cite{book}. If $C {\
\lower-1.2pt\vbox{\hbox{\rlap{$>$}\lower5pt\vbox{\hbox{$\sim$}}}}\ } 3.1$, i.e.
if $\phi_0 {\
\lower-1.2pt\vbox{\hbox{\rlap{$>$}\lower5pt\vbox{\hbox{$\sim$}}}}\ } 3.1M_{\rm P}
\sim 3.7\times 10^{19}$ GeV, the
universe expands  more than $e^{60}$ times, and  it becomes very flat.
Meanwhile, for $C \ll 3$ the universe always remains ``underinflated'' and very
curved, with $\Omega \gg 1$. We emphasize again that in this particular model
``underinflation" does not lead to any problems with homogeneity and isotropy.
The only problem with this model is that in order
to obtain $\Omega$ in the interval between $1$ and $2$ at the present time one
should have the constant $C$ to be fixed somewhere near $C = 3$ with an
accuracy of few percent. This is a fine-tuning, which does not sound very
attractive. However, it is important to realize that we are not talking about
an exponentially good precision; accuracy of few percent is good enough.

A similar result can be obtained even without changing the shape of the
effective potential. It is enough to assume that the field $\phi$ has a
nonminimal interaction with gravity of the form $-{\xi\over 2} R\phi^2$.  In
this case inflation becomes impossible for $\phi > {M_{\rm P}\over \sqrt{8\pi\xi}}$
\cite{Maeda}. Thus in order to ensure that only closed inflationary
universes can be produced during the process of quantum creation of the
universe in the theory ${m^2\over 2} \phi^2$ it is enough to assume that
${M_{\rm P}\over \sqrt{8\pi\xi}} < 3.1 M_{\rm P}$. This gives a condition $\xi > {1\over 72\pi} \sim
4\times10^{-4}$.

There is at least one serious problem with this scenario. Eq. (\ref{2}) suggests  that it is much
more natural for the universe to be created with the density very close to the
Planck density. However, the effective potential (\ref{1}) at the Planck
density is extremely steep.  Therefore such a universe   will not typically
enter the
inflationary regime, and will collapse within a very short time. This suggests that most of the universes described by the model (\ref{1}) will not inflate at all, or will inflate by a factor much less than $e^{60}$. In other words, most of the universes will be short living, and if they survive until the present time they will probably have very large $\Omega$.

This is the problem that we are going to study in our paper.

\section{Closed quasi-de Sitter universe}\label{sec:closed}

\begin{figure}\label{dS}
\centering\leavevmode\epsfysize=5.3cm \epsfbox{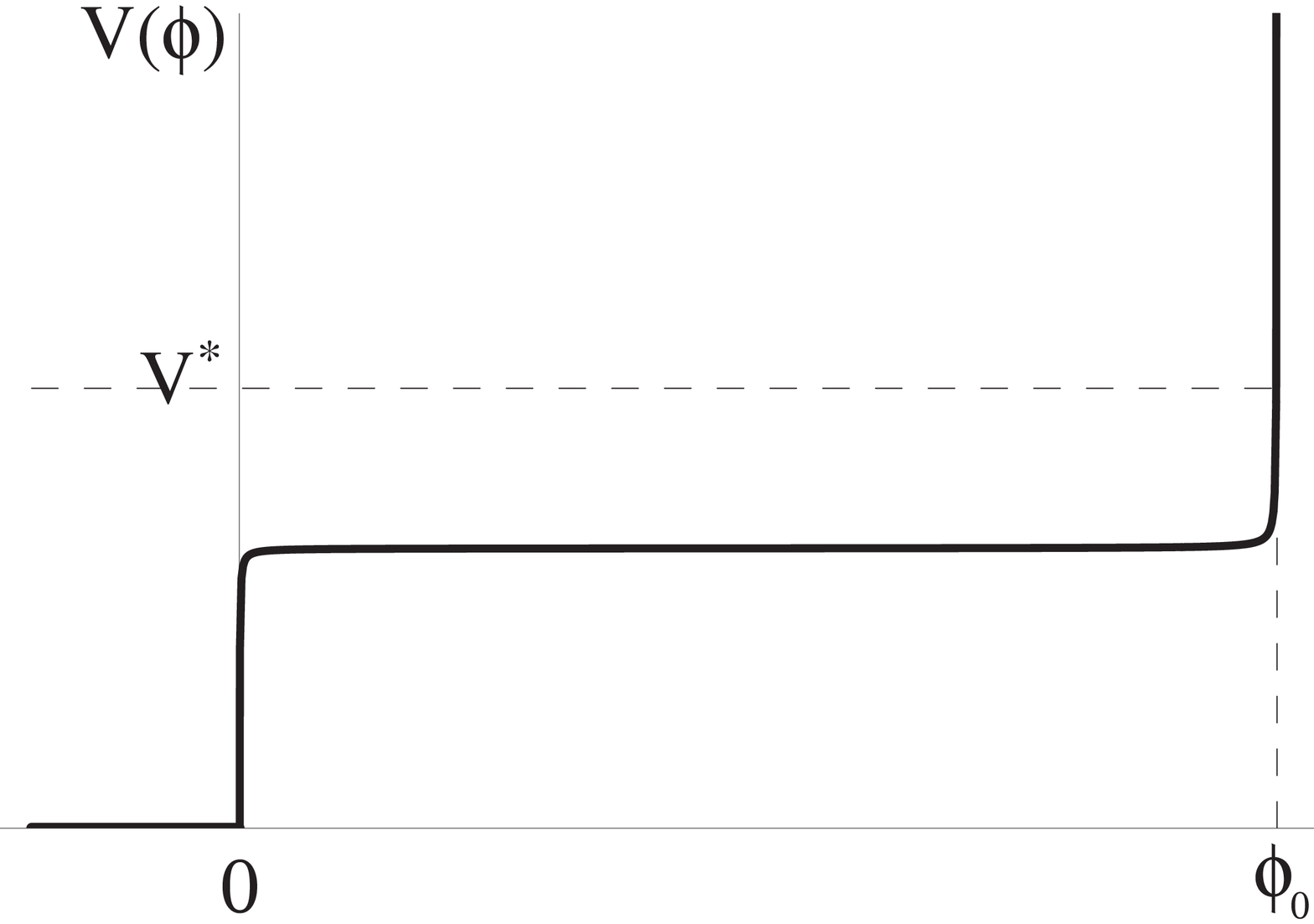}

\

\caption[Fig001]{\label{plot1a} Effective potential in our toy model; $V^* = {3\over 2} V$.}

\end{figure}

Consider for simplicity a toy model with the following step-like effective
potential: $V(\phi) = 0$ at $\phi < 0$; $V(\phi) = V = const$ at $0
<\phi<\phi_0$. We will also assume that the effective potential sharply
rises to indefinitely large values in a small vicinity of $\phi =
\phi_0$,  see Fig. \ref{plot1a}. This model is very well suited for the
description of certain versions of F-term hybrid inflation in
supergravity. In such models the effective potential is nearly constant
at small $\phi$ and it exponentially rises at large $\phi$. Inflation
ends suddenly at some small value of $\phi$, which in our model
corresponds to $\phi = 0$.

To study the evolution of a closed universe filled by a homogeneous
scalar field in this scenario one should solve the set of two equations,
in units $M_p = 1$.
\begin{eqnarray}\label{1u}
&&\ddot\phi +3{\dot a\over a}\, \dot\phi = -V'(\phi) \ , \nonumber\\
&&\ddot a = {8\pi\over 3}\, a\, (V -\dot\phi^2) \ .
\end{eqnarray}
Suppose that the (closed) universe described by this model appeared ``from
nothing'' in a state with the field $\phi \geq \phi_0$ at the point with
$\dot a = 0$, $\dot\phi = 0$ and the potential energy density $V_0 \geq V$. It will be convenient to represent $V_0$ as  $V^* -\Delta V$, where $V^*= {3\over 2}V$ see Fig. \ref{plot1a}. If the effective
potential at $\phi > \phi_0$ grows very sharply, then the field
instantly falls down to the plateau with $V(\phi) = V$, and its potential
energy becomes converted to the kinetic energy, ${1\over 2} \dot\phi^2 =
{V\over 2}-\Delta V$, so that $\dot \phi = -\sqrt{V-2\Delta V}$. Since this happens
nearly instantly, at that time one still has $\dot a = 0$. Thus, to study
inflation in this scenario one should solve  equations (\ref{1u}) for
$V(\phi) = V = const$ in the interval $0 <\phi<\phi_0$ with initial
conditions $\dot a = 0$, $\dot \phi = -\sqrt{V-2\Delta V}$.

These equations have a very interesting solution in the particular case
when $\Delta V = 0$. In this case the term $V -\dot\phi^2$ vanishes, the
acceleration of the scale factor $\ddot a$ vanishes too, and since
initially $\dot a = 0$, the universe remains static. According to our
results concerning creation of the universe, its initial size is equal to
$a = H_0^{-1}$, where $H_0 = \sqrt{8\pi V_0\over 3} = 2\sqrt{\pi
V}$. Thus, instead of inflating, the universe keeps its original size,
whereas the scalar field $\phi$ moves with the constant speed $\dot\phi
=\sqrt{V}$.

Obviously, this is a threshold regime.  For $\Delta V < 0$ (the 
initial energy density above $V^*$) one has $V
-\dot\phi^2< 0$, which means that $\ddot a < 0$. Thus the universe starts
moving with negative acceleration from the state $\dot a = 0$. As a
result, the term $3{\dot a\over a}$ becomes negative. This corresponds to
negative friction, which makes the motion of the field $\phi$ even faster
and the term $V -\dot\phi^2$ even more negative. Such a universe rapidly
collapses.

On the other hand, for $\Delta V > 0$ (initial energy below $V^*$) one has $V -\dot\phi^2> 0$, $\ddot a > 0$ and $3{\dot a\over a}> 0$. The value of $\dot \phi $
rapidly decreases, and the universe enters inflationary regime.

We will introduce here a small time-dependent function 
\begin{equation}\label{beta}
\beta(t)= {V-\dot\phi^2\over 2 V}\ll 1 \ .
\end{equation} 
The beginning of the stage of
inflation may be delayed for indefinitely long time if the initial value  
$\beta_0 = {\Delta V\over   V} $ is sufficiently small. 
But once expansion of the universe begins
in earnest, it very rapidly approaches the simple exponential regime $a
\sim e^{Ht}$. Indeed, in the regime with $V = const$ the scalar field
satisfies equation
\begin{equation}\label{2u}
\ddot\phi +3{\dot a\over a}\, \dot\phi =0 \ ,
\end{equation}
which implies that 
\begin{equation}\label{3u}
\dot\phi(t)  = \dot\phi_0 \ \left({a_0\over a(t)}\right)^3 \ ,
\end{equation}
Here $\dot\phi_0$ is the initial velocity of the field $\phi$ immediately after it rolls down to the plateau $V(\phi) = const$. This relation means, in particular, that once the size of the universe grows 2 times, the kinetic energy of the scalar field drops down
 64 times, and the universe begins expanding as $a = e^{Ht}$ with $H=\sqrt{8\pi V\over 3}$. 

The resulting equation for the scale factor reads:
\begin{equation}\label{4u}
{\ddot a } = {8\pi\over 3}\, a(t)\, \left(V -{\dot\phi_0^2 a_0^6\over a^6(t)}\right) \equiv {16\pi\over 3}\, a\, \beta(t) \ V  \ ,
\end{equation}
This equation can be solved numerically, but here we would like to make a simple qualitative analysis of its solutions for $\beta(0) \equiv \beta_0 \ll 1$, where \begin{equation}\label{4a}
\beta_0 = {\Delta V\over   V} =  {1\over 2} 
\left(1 -{\dot\phi_0^2\over V}\right) \ .
\end{equation}

First of all, let us notice that after $a(t)$ grows and inflationary regime settles up, the scalar field gradually stops moving. Indeed, according to  Eq. (\ref{4u}), $\dot \phi = \dot\phi_0 e^{-3Ht}$, and therefore the field $\phi$ after the beginning of inflation can move only at a finite distance 
\begin{equation}\label{5u}
\Delta\phi_{\rm inf} =  { \dot\phi_0\over 3H} \approx  -{1\over 2\sqrt{6\pi}}\ ,
\end{equation}
in Planck units $M_p = 1$.

For small positive $\beta$ the whole process can be approximately divided
into two stages. In the beginning of the process, when $a \approx a_0$ and
$\beta(t) \approx \beta_0$, equation
\begin{equation}
\ddot a = {16\pi\over 3}\, a\, \beta V
\end{equation}
 yields
\begin{equation}
a = a_0  \left(1+  {8\pi \beta(0) V\over 3}\, t^2\right) \ .
\end{equation}
 Using the relation $\dot\phi^2 \sim a^{-6}$ one
finds that the value of $\beta$ becomes two times greater than $\beta(0)$
within the time
\begin{equation}\label{t1}
\Delta t_1 \approx {1\over 2\sqrt{3\pi V }} \ .
\end{equation}
During this time the field $\phi$ decreases by
\begin{equation}\label{phi1}
\Delta\phi_1\sim  \dot \phi(0)  \Delta t_1 \approx -{1\over 2\sqrt{3\pi}} \ ,
\end{equation}
which is similar to $\Delta\phi_{\rm inf}$, Eq. (\ref{5u}).
 After that the rate of growth of the scale factor doubles, but $a(t)$ remains practically unchanged (for $\beta_0 \ll 1$). This means that the whole process effectively gets a fresh start. After the time $\Delta t_2 \approx \Delta  t_1$, which does not depend on $\beta$, the field $\phi $ decreases by $\Delta\phi_2 \approx \Delta\phi_1$, and the rate of growth of $a(t)$ doubles again. This process continues until $\beta(t)$ approaches $1/2$. Since at each interval $\Delta t_i$ the value of $\beta$ doubles, the number of the intervals after which $\beta(t)$ approaches $1/2$ is proportional to $\ln \beta_0$. Then the universe enters inflationary regime, the field  moves last time, by $\Delta\phi_{\rm inf}$,  and  the field $\phi$ finally stops.  

These conclusions are in a good qualitative agreement with the results of a numerical investigation (for $\beta_0 \ll 1$), which shows that 
inflation begins when the field $\phi$ approaches $\phi_{\rm inf}$, where
\begin{equation}\label{phib}
\phi_{\rm inf} \approx \phi_0 + 0.1 +0.15\  \ln \beta_0 \ .
\end{equation}

\begin{figure}
\centering\leavevmode\epsfysize=4.5cm \epsfbox{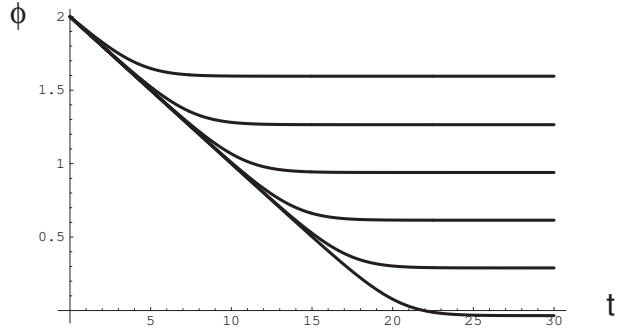}

\

\caption[Fig001]{\label{deSitt}  Behavior of the scalar field in our toy model. If the field starts its motion with sufficiently small velocity, inflation begins immediately. If it starts with large initial velocity $\dot\phi_0$, due to falling from the large `height' $V(\phi) = V^* -\Delta V$, the universe never inflates. Inflation begins at $\phi \approx \phi_0 +0.1 +0.15 \, \ln {\Delta V\over V}$.}
\end{figure}

Inflation takes place only in the interval between $\phi_{\rm inf} >0$ and $\phi = 0$. If $\phi_{\rm inf} < 0$, inflation does not happen at all, and the universe rapidly collapses.

Thus inflation happens if
\begin{equation}\label{phitot1}
\phi_0 > -0.1-    0.15\  \ln \beta_0 \ .
\end{equation}

Now let us study the implications of these results for the theory of quantum creation of a closed inflationary universe in this scenario. First of all, we will assume that, in accordance with \cite{Creation} - \cite{Vilenkin}, the probability of creation of a closed universe containing homogeneous field $\phi$ with the effective potential $V(\phi)$ is given by Eq. (\ref{2}),
\begin{equation}\label{TUNN1}
P \sim e^{- 2|S|} = \exp \left(-{3 M_p^4\over 8 V(\phi)}\right) \ .
\end{equation}
As we see, the greater $V(\phi)$, the more probable it is that the universe is created. However, all universes that are created with $V(\phi) \geq V^* = {3\over 2 } V $ immediately collapse without giving life any chance to appear. Thus the high probability of quantum creation of the universes with large $V(\phi)$ in this context does not mean much: For all practical purposes such universes simply do not exist. 

The possibility of creation of the universes with $V(\phi) < {3\over 2}\, V$ requires a much more thorough investigation. Indeed, it is  most probable for the universe to be created with the energy density very close to $V(\phi) = {3\over 2}\, V$, i.e.  with as small $\beta_0$ as possible. However, if $\beta_0$ is so small that $0.15\, |\ln \beta_0| > \phi_0$,  then, according to our results, the field $\phi$ reaches the point $\phi = 0$ and falls from the cliff before inflation begins. In this case the universe also remains small and immediately collapses.  If $\beta_0$ is small, but not  small enough, so that the stage of inflation is very short, we also have a problem. Indeed, in order to explain observational data we need to have approximately 60 e-folds of inflation (under the simplest assumptions about reheating).

Of course, one could argue, just as we did before, that the universes experiencing less than 60 e-folds of inflation  cannot support human life and therefore should be ignored. However, in this case the situation is more complicated.

Indeed, let us estimate first the conditional probability that  the universe is created with the energy density ${3\over 2} V - \beta_0 V$, under the condition that its energy density $V$ is smaller than  $V^* ={3\over 2} V$. (As we explained, all universes with $V>V^*$ immediately die and therefore must be discarded in the calculation of the probabilities.)
\begin{eqnarray}\label{TUNN2}
P &\sim& e^{- 2|S|} = \exp \left(-{3M_p^4  \over 8({3\over 2} -\beta_0) V} +{3M_p^4  \over 8 \cdot{3\over 2} V}\right)\nonumber \\ &\sim& \exp \left(-{M_p^4\beta_0 \over 6 V}\right) \ .
\end{eqnarray}
This expression tells that the probability of creation of the universe  with $\beta_0 \not = 0$ will be exponentially suppressed unless the universe is created very close to the threshold $V(\phi_0) = {3\over 2} V$, with $\beta_0 < {6V\over M_p^4}$.

Thus, the probability of creation of the universe is not suppressed for $\beta_0 < {6V\over M_p^4}$, which means that the probability of creation of an {\it inflationary} universe is not suppressed for 
\begin{equation}\label{phitot1a}
\phi_0 > -0.1-    0.15\  \ln {6V\over M_p^4}.
\end{equation}

As a particular example, one may consider the case $V \sim 10^{-11} M_p^4$, as in the simplest models of chaotic inflation at the end of inflation \cite{book}. In this case the probability of creation of an {\it inflationary} universe is not suppressed for $\beta_0 < {6\cdot 10^{-11}}.$ 
This means that if the distance between the initial point, where the effective potential blows up, and the final point, where the effective potential drops down to zero, is greater than
\begin{equation}\label{phitot1b}
\phi_0 > -0.1-    0.15\  \ln\, ({6\cdot 10^{-11}}) \sim 3.4 M_p\ ,
\end{equation}
the probability that the universe enters a stage of inflation is not exponentially suppressed.

We would like no note that our main argument was based on the simple fact that the inflationary regime is an attractor in the phase space of all possible trajectories. That is why the inflationary regime is reached very fast, unless   $\beta_0$ is exponentially small. But if  $\beta_0$ is exponentially small, the value of  ${\Delta V} = \beta_0 V$ is also exponentially small. As a result, the exponential suppression of the probability of the tunneling to the state   with the energy density $V^*-\Delta V$ as compared to the tunneling to the state with the energy density   $V^*$  disappears, and a long stage of inflation becomes possible. We expect this result to remain qualitatively correct even if the expression for the probability of tunneling to the non-inflationary part of the potential significantly differs  from Eq. (\ref{2}).

Let us consider two different possibilities. 

1) Suppose first that $V \sim 10^{-11} M_p^4$ and $\phi_0$ is a bit smaller than $3.4 M_p$, e.g. $\phi_0 = 3M_p$. Then inflation may occur in this scenario only if  $\beta_0 \gtrsim 2\times 10^{-9}$. The probability of creation of an inflationary universe with $\beta_0 \gtrsim 2\times 10^{-9}$ will be suppressed by $e^{-200}$. One could argue that this is not too bad because all non-inflationary universes immediately collapse. The real problem, however, is that in this case the probability to obtain the universe that inflates $e^{59}$ times will be much greater than the probability to obtain the universe that expands $e^{60}$ times. This means that the universe will have as large value of $\Omega$ as possible,  compatible with the existence of our life. This does not allow us to explain why do we live in the universe with $\Omega < 1.1$.

2) Suppose now that $\phi_0$ is considerably greater than $3.4 M_p$, e.g. $\phi_0 = 10M_p$. In this case  inflation  typically begins at $\phi \sim 6.6 M_P$. After the beginning of inflation, the field $\phi$ stops moving when it passes the distance $|\Delta\phi_{\rm inf}| =   {1\over 2\sqrt{6\pi}} \sim 0.114$. If it stops before it reaches $\phi = 0$, inflation continues forever, so the universe will be  unsuitable for life as we know it.

Thus, the conclusions for the simplest model are not very optimistic. The main reason for it is the constancy of the potential in the interval from $\phi_0$ to $\phi = 0$. This property of the potential implies that the universe typically either collapses very soon, or inflates forever. As we will see, this particular problem disappears in the realistic inflationary models.

\section{Chaotic inflation with $V = m^2\phi^2/2$}

Now we are going to study realistic models of chaotic inflation. 

\begin{figure}
\centering \leavevmode\epsfysize=5.5cm \epsfbox{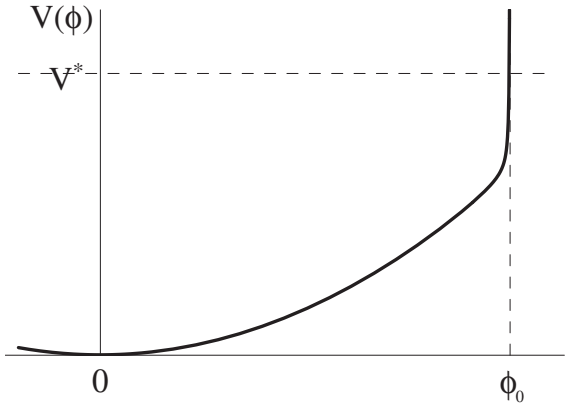}

\
 
\caption[Fig001]{\label{fig2} { Effective potential in the model with the potential that is equal to  $m^2\phi^2/2$ at $\phi < \phi_0$ and blows up at $\phi > \phi_0$.}}
\end{figure}

Consider for example the theory with a potential which is given by $V = m^2\phi^2/2$ for $\phi < \phi_0$, and  becomes extremely steep at $\phi > \phi_0$, see Fig. \ref{fig2}.  If the universe is created at $\phi$ just above $\phi_0$, in the point with $V_0 > V(\phi_0) = m^2\phi_0^2/2$, the field immediately falls down to $\phi_0$ and acquires velocity given by $(\dot\phi_0)^2/2 = V_0 - V(\phi_0)$. Just as before, if the field starts its motion with sufficiently small velocity, inflation begins immediately. If it starts with large initial velocity $\dot\phi_0$, the universe never inflates.

However, there is a slight difference here. In the toy model discussed above the boundary between inflation and no inflation was independent of $\phi_0$, and it was given by $V_0-V(\phi_0) = V(\phi_0)/2$. In the realistic models this boundary does not differ much from $V_0-V(\phi_0) = V(\phi_0)/2$, but the difference does exist and it depends on $\phi_0$.

We will denote $V^*(\phi_0)$ the critical initial value of $V$, such that inflation still exists for the field $\phi$ falling to the point $\phi_0$ from the height $V_0 < V^*$, and disappears at $V_0 > V^*$. Then, just as in the previous section, we introduce the parameter $\beta_0 = \Delta V/V(\phi_0)$, where $\Delta V = V^* - V_0$ and $V(\phi_0) = m^2\phi_0^2/2$.

As an example, we considered a model where the effective potential blows up at $\phi > \phi_0 =O(10)$ (i.e. at $\phi > O(10) M_p$). In this case we have found that   inflation begins when the field $\phi$ rolls down to the point 
\begin{equation}\label{phi2beg}
\phi_{\rm inf}\approx \phi_0 - 0.05 + 0.15 \, \ln\,\beta_0.
\end{equation}
  This equation becomes invalid for $\phi \ll 1$, in which case inflation never happens. Interestingly, this equation is very similar to the equation (\ref{phib}) for the toy model considered in the previous section. This indicates that our results are not very sensitive to the choice of a particular inflationary model.

During inflation in this scenario the universe expands $e^{2\pi\phi^2}$ times, so one needs to have $\phi \approx 3.1$ to achieve $60$ e-folds of inflation. 
Now let us remember that for $V(\phi_0) \sim 10^{-11} M_p^4$ the probability of creation of universes with $\beta_0 < {6\cdot 10^{-11}}$ is not exponentially suppressed. The probability to start with $\beta_0 \ll {6\cdot 10^{-11}}$ is suppressed because of the small phase space corresponding to these values of $\beta_0$. Therefore one can argue that it is most probable to have $\beta_0\sim 6\cdot 10^{-11}$. This means that inflation typically starts at 
\begin{equation}\label{phi2beg3}
\phi_{\rm inf}\approx \phi_0 - 3.5 \ .
\end{equation}

\begin{figure}
\centering\leavevmode\epsfysize=4.5cm \epsfbox{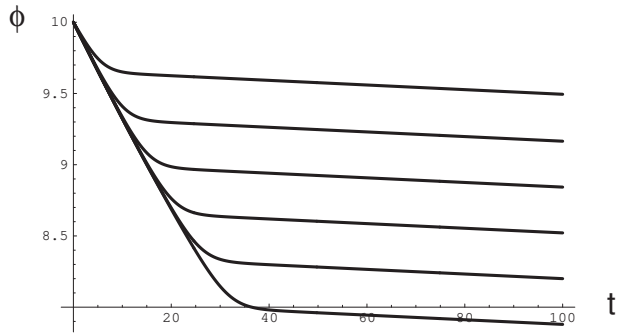}
\caption[Fig001]{\label{onefielddistrib} Behavior of the scalar field in the theory $m^2\phi^2/2$. If the field starts its motion with sufficiently small velocity, inflation begins immediately. If it starts with large initial velocity $\dot\phi_0$, due to falling from the large value of $V(\phi)$, the universe never inflates. Inflation begins at $\phi \approx \phi_0 - 0.05 +0.15 \, \ln {\Delta V\over V}$.}
\end{figure}

The rest is obvious. For $\phi_0 = 10$, inflation starts at $\phi_{\rm inf} \sim 6.5$, the universe inflates $e^{265}$ times and becomes absolutely flat.
The universe inflates $e^{60}$ times for $\phi_{\rm inf}= 3.1$, i.e. for $\phi_0 = 3.1+3.5 = 6.6$. If this leads to $\Omega = 1.1$, then an increase of $\phi_0$ by $1\%$ increases the number of e-folds by $O(1)$, which makes  the universe almost exactly flat.

If this is correct, tuning of the value of $\Omega$ to be in the range $\Omega < 1.1$ requires fine-tuning of the parameter $\phi_0$ with accuracy about $1\%$. Whereas this fine-tuning is not very attractive, it is certainly not as outrageous as the fine-tuning required in all existing models of dark energy.

\section{Discussion}
In this paper we discussed a model of a closed inflationary universe based on a  modified version  of the simplest chaotic inflation scenario. If the main idea is correct, one can implement it in a variety of different models. For example, one can consider a potential with a flat maximum, like in new inflation, and add to it a high `tower' on the top. The universe will be created at the top of the `tower,'  but it will be unable to stay there. When it falls down, it  rolls downhill with a large initial velocity, and only gradually approaches the asymptotic inflationary regime. As a result, the stage of inflation can be made short, and the large-scale density perturbations will be suppressed. Other models of a similar type can be constructed in the context of hybrid inflation \cite{Hybrid}.

Similar considerations may apply to other types of compact universes. For example, one of the first papers on quantum creation of the universe ``from nothing"  was devoted to the possibility of a quantum creation of a flat universe with nontrivial topology \cite{ZelStar}. Some of our arguments should be modified in this case. In particular, in certain cases the probability of quantum creation of the universe will not be suppressed at all, whereas in some other cases the exponential suppression will occur due to the Casimir effect, and it will not depend on $V(\phi)$ \cite{ZelStar}. 

The reason for this effect is very simple. Unlike a closed inflationary universe, flat universe can emerge from the singularity without any need to tunnel through the barrier. The only barrier that could appear is due to the finite size of the universe, which alters the spectrum of quantum fluctuations there (Casimir effect). However, this effect appears at the one-loop level only, and it may be strongly suppressed in supersymmetric theories.

Therefore one may even argue that among all possible universes created ``from nothing'', the flat universes with nontrivial topology are the first to emerge, so it would be more natural for us to live in one of those rather than in the closed universe. One could turn this argument another way around and argue that if we do not find the specific anisotropy of the CMB associated with the periodicity of the universe, this will make is even less likely that our universe is closed. This strengthens the standard inflationary prediction $\Omega =1$.

The easiest way to obtain a short stage of inflation in all of these models (closed universe or flat universe with nontrivial topology) is to make sure that inflation cannot be too long. In the models discussed above, this was realized by making the potential very curved at large $\phi$. This leads to a combination of two different effects. First of all, at the onset of inflation the curvature of the potential was very large, so the long-wavelength perturbations of the scalar field were suppressed. Secondly, the speed of the rolling field was very large. Both of these effects tend to suppress the density perturbations produced at the very early stages of inflation. This effect may account for the suppression of the large-angle CMB anisotropy observed by WMAP. To check the validity of our conjecture with respect to the closed universe case one should find the corresponding `gondola' instantons \cite{Bousso:1998ed} and then use, e.g., the methods of \cite{Gratton:2001gw}, applying them to our potential rapidly growing at large $\phi$. For the flat universe case one should make a similar investigation based on \cite{ZelStar}. The effect of suppression of the large-angle CMB anisotropy may be more pronounced in the theories of the hybrid inflation type, or in new inflation, where the possible large-scale contribution of gravitational waves will be suppressed.

 Note, however, that if our only goal is to suppress large-scale perturbations, there are many other ways to do so \cite{CKLP}. For example, the large-scale perturbations are strongly suppressed in the flat universe case in the simplest version of the F-term hybrid inflation \cite{Linde:1997sj}.

All of the proposed models describing a closed (compact) inflationary universe are artificial and fine-tuned. Therefore the flatness of the universe and the flatness of the spectrum of density perturbations remain a generic prediction of a broad class of inflationary models. However, it is good to know that, if needed, we can describe compact universes without giving up all  advantages of inflationary
cosmology. The fine-tuning required for the construction of the inflationary models of a closed (compact)  universe is much smaller and much easier to achieve than the incredible fine-tuning required for the explanation of the enormously large mass and entropy of the universe,  of its homogeneity and isotropy, and of the observed  anisotropy of CMB without using inflation.  

It is a pleasure to thank  C. Contaldi,
L. Kofman, V. Mukhanov, and M. Peloso  for
useful discussions.  This work was supported in part
by NSF grant PHY-987011 and by the Templeton Foundation grant
No. 938-COS273.

\end{document}